\documentclass[prl,twocolumn,groupedaddress]{revtex4}
\usepackage{graphics,epsfig}

\begin{document}

%\draft

\title{Effect of congestion costs on shortest paths through complex networks}
\author{Douglas J. Ashton, Timothy C. Jarrett, Neil F. Johnson} \address{Physics
  Department, Oxford University, Oxford OX1 3PU, U.K.}

\date{\today}

\begin{abstract} We analyze analytically the effect of congestion costs within a
physically relevant, yet exactly solvable network model featuring central hubs.
These costs lead to a competition between centralized and decentralized
transport pathways. In stark contrast to conventional no-cost networks, there
now exists an optimal number of connections to the central hub in order to
minimize the shortest path. Our results shed light on an open problem in
biology, informatics and sociology, concerning the extent to which decentralized
versus centralized design benefits real-world complex networks.

\end{abstract}

\maketitle

\newpage

The interplay between structure and function in complex networks, has become a major research
topic in physics, biology, informatics, and sociology
\cite{watts98,nets,nets2,charges,central, search, gradient}. For example, the very same links,
nodes and hubs that help create short-cuts in space for transport, may become congested due to
increased traffic yielding an increase in
transit time
\cite{charges}. Unfortunately there are very few analytic
results available concerning network congestion and optimal pathways in real-world networks
\cite{charges,central,search,gradient}.

In this paper, we provide exact analytic results for the effects of congestion costs in networks with
a combined ring-and-star topology. Figure 1(a) shows an example of our model network with
$N=1$ central hub.  In addition to the fact that it is analytically tractable and posseses a
topology which is distinct from Refs.
\cite{charges,central,search,gradient}, our model network is of direct relevance
to a wide range of biological,
  computational and socio-economic systems in which there is a potentially congested
  central node(s). Figure 1(b) shows the nutrient transport in a
laboratory-grown fungus \cite{appl}. The major transport pathways pass through a central hub (i.e.
centralized transport) with some minor pathways around it (i.e. decentralized transport). It is
an important yet open question in biology as to how organisms such as fungi make a trade-off between
centralized and decentralized transport, communication and control.  A related scenario with a
similar topology, concerns the new congestion charge scheme in London which aims to dissuade drivers
from passing through the central zone. Airlines must balance the costs and benefits of stopovers at
major, yet potentially overcrowded, airport hubs. Similar trade-offs between centralized and
decentralized routing, communication and control arise in data networks, manufacturing
supply-chains, and government. Even for crime or terrorist networks, one can ask how the Mafia's
approach of passing all decisions through a central `Godfather' compares to the apparently headless
form of modern terrorist cells.  More generally, our model network could be used to describe
clusters or motifs within larger networks in which relatively isolated hubs are connected to
lower-connectivity nodes (e.g. scale-free network).

%%%%%%%%%%%%%%%%%%%%%%%%%% FIGURE 1 %%%%%%%%%%%%%%%%%%%
\begin{figure}[ht]
\includegraphics[width=.35\textwidth]{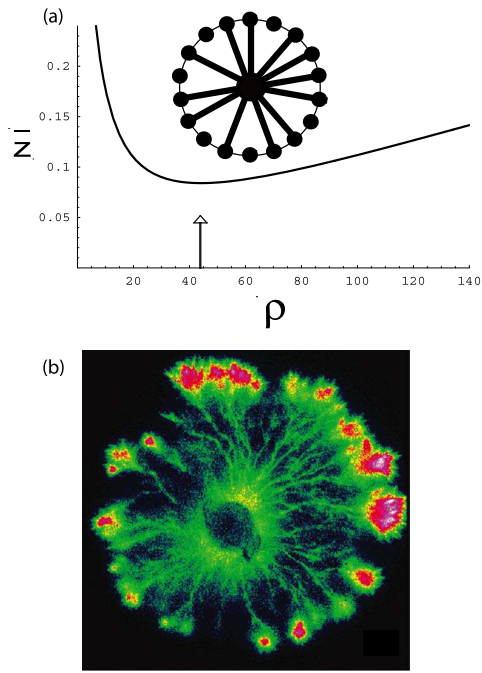}
\caption{(Color online) (a) Our model network showing transport pathways through the
central hub (connections of length $1/2$ denoted by thick lines) and around the ring (connections of
length
$1$ denoted by thin lines). Graph shows average shortest path length between any two nodes in a
$n=1000$ node ring,
  with a cost-per-connection to the hub of $k=1$. There is an optimal
  value for the number of connections ($\rho\equiv p n\approx 44$) such that the average
  shortest path length $\bar{\ell}$ is a minimum.  We denote this minimal
  shortest path length as $\label{single lbar def} \bar{\ell}\equiv\bar{\ell}|_{\rm
    min}$.
(b) Photon scintillation image showing the nutrient distribution 
within a laboratory-grown fungus  {\em Phanerochaete velutina}. Nutrient
density increases going from blue to green to red.}
\label{fig:figure1}
\end{figure}
%%%%%%%%%%%%%%%%%%%%%%%%%% FIGURE 1 %%%%%%%%%%%%%%%%%%%

Our model represents a generalization of Ref. \cite{DM} to the case of non-zero congestion
costs. Each of the $n$ nodes around the ring is connected to its nearest neighbors by
a link of unit length. These links are directed in the `directed' model, and undirected in the
`undirected' model. With a probability $p$ any node can be
attached to the central hub by a link of length $\frac{1}{2}$. The links to the hub are
always undirected. For both the directed and undirected models, explicit
expressions can be derived for  the probability $P(\ell,m)$ that the shortest
path between any two nodes on the ring is $\ell$, given that they are
separated around the ring by length $m$.  Summing over all $m$ for a given
$\ell$ and dividing by $(n-1)$ yields the probability $P(\ell)$ that the shortest path between two
randomly selected nodes is of length $\ell$. The average value for the shortest
path across the network is then $\label{single lbar def} \bar{\ell}=\sum_{\ell
  =1}^{n-1}\ell P(\ell)$.  For the undirected model,  the expressions are
  more cumbersome because there are more paths with the same length.  However,
  defining $n P(\ell)\equiv Q(z,\rho)$ where $\rho\equiv p n$ and $z\equiv
  \ell/n$, there is a simple relationship between the undirected and directed 
  models in the limit $n\rightarrow \infty$ with $p \rightarrow 0$, i.e.
  $Q_{undir}(z,\rho)=2Q_{dir}(2z,\rho)$ \cite{DM}.  The models only differ in
  this limit by a factor of two: $z \rightarrow 2z$, with $z$ now running from
  $0$ to $1/2$. The results which follow were obtained by generalizing this
  procedure.  

%%%%%%%%%%%%%%%%%%%%%%%%%% FIGURE 2 %%%%%%%%%%%%%%%%%%%
\begin{figure}[ht]
\includegraphics[width=.35\textwidth]{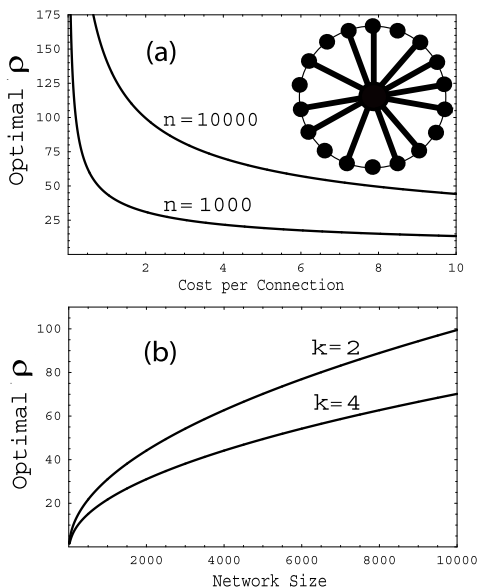}
\caption{Minimal shortest path length $\bar{\ell}|_{\rm min}$
(i.e. minimum value of $\bar{\ell}$) as obtained from Eq. (\ref{cost min lbar
                                                           expanded}).  (a)
  Optimal number of connections $\rho\equiv p n$ as a function of the cost-per-connection  $k$ to the
hub.
  Results are shown for $n=1000$ and $n=10000$. (b) Optimal number
  of connections $\rho$ as a function of the network size. Results are shown for
  $k=2$ and $k=4$.}
\label{fig:figure2}
\end{figure}
%%%%%%%%%%%%%%%%%%%%%%%%%% FIGURE 2 %%%%%%%%%%%%%%%%%%%

We add a cost $c$ every time a path passes through the central hub.  This cost
$c$ is expressed as an additional path-length, however it could also be
expressed as a time delay or reduction in flow-rate for transport and
supply-chain problems.  We consider three cases: (1) constant cost $c$ where $c$
is independent of how many connections the hub already has, i.e.  $c$ is
independent of how `busy' the hub is; (2) linear cost $c$ where $c$ grows
linearly  with the number of connections to the hub, and hence varies as $\rho
\equiv np$; (3) nonlinear cost $c$ where $c$ grows with the number of pairs
connected directly across the network, and hence varies as $\rho^2$.

For a general, non-zero cost $c$ that is independent of $\ell$ and $m$, we can
write (for a network with directed links): {\small \begin{eqnarray} P(\ell ,
    \ell \leq c) &=& \frac{1}{n-1} \\ P(\ell < m, \ell > c) &=&
      (\ell-c)p^2(1-p)^{\ell-c-1} \label{p-ell-lt-m} \\ P(\ell = m, \ell > c) &=&
      1-p^2\sum_{i-c=1}^{\ell-c-1}(i-c)(1-p)^{(i-c)-1} \end{eqnarray}} \noindent
      Performing the summation gives: \begin{equation} P(\ell = m, \ell > c) =
      (1+(\ell-c-1)p)(1-p)^{\ell-c-1} \end{equation} The shortest path distribution is
      hence: 
{\small \begin{displaymath} 
P(\ell) = \left\{ \begin{array}{ll}
\frac{1}{n-1} & \ \ \textrm{$\forall$ $\ell \leq c$}\\
\frac{1}{n-1}\bigl[1+(\ell -c-1) p \\
\ \ \ \ \ \ + (n-1-\ell )(\ell -c)p^2 \bigr](1-p)^{\ell-c-1} & 
\ \ \textrm{$\forall$  $\ell > c$}
\end{array} \right.  
\end{displaymath}}
Using the same analysis for undirected links yields a simple relationship
between the directed and undirected models.  Introducing the variable $\gamma
\equiv \frac{c}{n}$ with $z$ and $\rho$ as before, we may define $nP(\ell)
  \equiv Q(z, \gamma, \rho)$ and hence find in the limit $p \rightarrow 0$, $n
  \rightarrow \infty$ that $Q_{undir}(z,\gamma,\rho)=2Q_{dir}(2z,
      2\gamma,\rho)$.  For a fixed cost, not dependent on network size or the
  connectivity, this analysis is straightforward.  Paths of length $l \leq c$
  are prevented from using the central hub, while for $l>c$ the distribution
  $P(l)$ is similar to that of Ref. \cite{DM}.

For linear costs, dependent on network size and connectivity and for $N=1$
central hub, we can show that there exists a {\em minimum} value of the average
shortest path $\label{single lbar def} \bar{\ell}$ as a function of the
connectivity to the central hub. Hence there is an {\em optimal} number of
connections to the central hub, in order to create the  minimum possible average
shortest path. We denote this minimal path length as $\label{single lbar def}
\bar{\ell}\equiv\bar{\ell}|_{\rm min}$.  Such a minimum is in stark contrast to
the case of zero cost per connection, where the value of $\bar{\ell}$ would just
decrease monotonically towards one with an increasing number of connections to the hub.
We now calculate the average shortest path, $\bar{\ell} = \sum_{\ell=1}^{n-1}
\ell P(\ell)$, which yields: 
{\small \begin{eqnarray}\label{cost min lbar expanded}
\bar{\ell }=\frac{(1-p)^{n-c}\bigl[3+(n-2-c)p\bigr]}{p^2(n-1)}
\ \ \ \ \ \ \ \ \ \ \ \ \ \ \ \ \ \ \ \ \ \ \ \nonumber\\
+\frac{p\bigl[2-2c+2n-(c-1)(c-n)p\bigr]-3}{p^2(n-1)}+\frac{c(c-1)}{2(n-1)}\ .
\end{eqnarray}} 
\newline  Figure 1 shows the functional form of $\bar{z} \equiv
\frac{\bar{\ell}}{n}$ with a cost of $1$ unit path-length per connection to the hub (i.e.
    $c=knp=k\rho$, with $k=1$).  The optimal number of connections in order that
$\bar{\ell}$ is a minimum is approximately $44$ and depends on $n$.  The
corresponding minimal shortest path $\bar{\ell}|_{\rm min}$ is approximately 85. An analytic
expression for $\bar{\ell}|_{\rm min}$ can be obtained by setting the differential of Eq. (\ref{cost
min lbar expanded}) equal to zero.  If $n$ is very large, one can
introduce a higher cost without compromising the minimal shortest path
$\bar{\ell}|_{\rm min}$ since in general the nodes are already much further from one another. We can
also investigate how many connections we should make for a given cost and network size, in order
to achieve the minimum possible shortest path $\bar{\ell}|_{\rm min}$.  This is
obtained by setting the differential of Eq.  (\ref{cost min lbar expanded}) equal to zero and
solving for $p$.  Figure 2(a) shows analytic results for the optimal number of
connections which yield the minimal shortest path $\bar{\ell}|_{\rm min}$, as a
function of the cost per connection for a fixed network size. Figure 2(b) shows
analytic results for the optimal number of connections which yield the minimal shortest path
$\bar{\ell}|_{\rm min}$, as a function of the network size for a fixed cost per
connection to the hub.

To gain insight into the underlying physics, we now make some approximations to
the exact analytic expressions.  For large $n$,
  or more importantly large $n-c$, the term $(1-p)^{n-c}\rightarrow e^{-\rho}$
  in Eq. (\ref{cost min lbar expanded}).  Provided that the cost per connection to the hub
  is not too high, the region containing the minimal shortest path
  $\bar{\ell}|_{\rm min}$ will be at a reasonably high $\rho$ (recall Fig. 1(a)).
  Hence we can neglect the exponential term and differentiate to find the
  minimum value of $\bar{\ell }$ with $c=knp=k\rho$.  It is reasonable to assume
  that at fixed $k$, optimal $\rho$ will increase with $n$ like $n^x$  where $0 < x \leq
  1$. In particular, one obtains diffusive behavior whereby $x\sim 1/2$.
  Specifically, $\rho \approx \sqrt{\frac{2n}{k}}$. For a large network (i.e.
      large $n$), we have therefore obtained a simple relationship between the
  number of  connections one should introduce in order to create the minimal
  average shortest path between any two nodes in the network, and the cost per
  connection to the hub.  It can be shown by comparing to Figure 2, that this analytic
  scaling relation is accurate even down to $n\sim 10$, but is particularly good
  for $n$ larger than $10^3$.

%%%%%%%%%%%%%%%%%%%%%%%%%% FIGURE 3 %%%%%%%%%%%%%%%%%%%
\begin{figure}[ht]
\includegraphics[width=.35\textwidth]{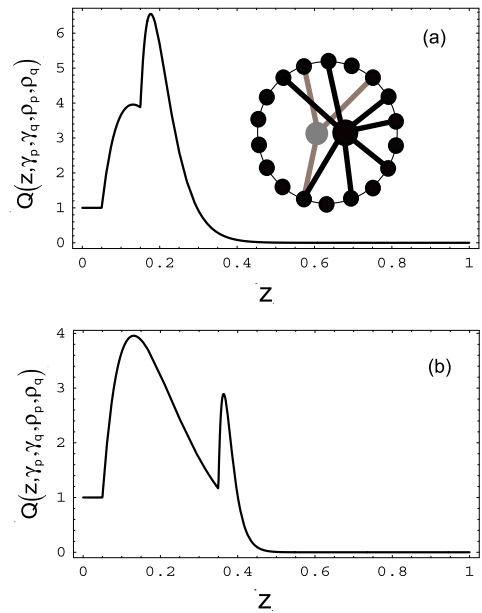}
\caption{Examples of the scaled probability distribution for a
network with $N=2$ hubs, where the two hubs have associated costs for travelling
through them.  In (a), $\rho_p=20$ and $\rho_q=10$ while the costs are
$c_p=0.15$ and $c_q=0.05$.  In (b),  $\rho_p=50$ and $\rho_q=10$ while the costs
are $c_p=0.35$ and $c_q=0.05$.}
\label{fig:figure3}
\end{figure}
%%%%%%%%%%%%%%%%%%%%%%%%%% FIGURE 3 %%%%%%%%%%%%%%%%%%%

Now we briefly turn to consider a specific yet physically reasonable example of
non-linear costs, in which the costs are taken to depend on the number of {\em
  pairs} which are connected via the hub.  In particular, we use $c=k(np)^2$.
  We obtain the analytic relationship $\rho \approx \sqrt[3]{\frac{n}{k}}$ which
  is the non-linear equivalent of the above result. Obviously, more accurate
  expressions can be obtained since we know the complete form of the analytic
  solution -- however these are too cumbersome algebraically to be presented
  here.

For linear costs, the lowest value of $\bar{\ell}$ one can achieve is $
\bar{\ell}|_{\rm min} \approx \sqrt{8kn}$. Setting $n=10^3$ and $k=1$ gives
$\bar{\ell}|_{\rm min}=89.4$, which agrees well with the exact analytic result shown in
Fig.  1. For non-linear costs, the minimal shortest path $\bar{\ell}|_{\rm min}
\approx \sqrt[3]{27kn^2}$.  These last results show that the minimal shortest
path $\bar{\ell}|_{\rm min}$ across the network grows like $n^\frac{1}{2}$ when
we impose linear costs while it grows
like $n^\frac{2}{3}$ when we put a cost on the number of direct connections
between nodes made via the hub (i.e. non-linear costs).  Corresponding results
for the undirected model can be easily obtained from the equations for the
directed model. For example for linear costs $c=knp$ and undirected links, we
obtain $\bar{\ell}|_{\rm min} \approx \sqrt{4kn}$ and $\rho \approx
\sqrt{\frac{n}{k}}$ for the minimal shortest path and the optimal connectivity.

The present analysis can be extended to multiple hubs, $N\geq 2$. For
simplicity, we focus here on the specific example of constant costs and $N=2$
(i.e. hub P, with nodes connected to it with probability $p$ and hub Q, with
 nodes connected to it with probability $q$) where the  cost associated with
each hub has value $c_p$ and $c_q$, with $c_p \geq c_q$.  The cost for using
{\em both} hubs is assumed to be infinite. It is not hard to imagine real-world
systems that employ multiple central hubs but which would not favour pathways
through more than one at a time (e.g. an airline passenger would avoid buying a
    ticket with two stop-overs).  Of course this
assumption may not always be realistic (see, for example, Ref. \cite{railway}).

We 
first consider what happens when $\ell > c_p \geq c_q$.  In this case, both hubs 
may be used and we may therefore write:
{\small
{\setlength\arraycolsep{1pt}
\begin{eqnarray}\label{thd-with-costs-before-sum}
P(\ell < m) &=& P_P(\ell < m, \ell > c_p)\bigl[ 1-
\sum_{i-c_q=1}^{\ell-c_q-1}P_Q(i < m, i > c_q)\bigr]\nonumber\\
  &+& P_Q(\ell < m, \ell > c_q)\bigl[ 1-
  \sum_{i-c_p=1}^{\ell-c_p-1}P_P(i < m, i > c_p)\bigr]\nonumber\\
  &-& P_P(\ell < m, \ell > c_p)P_Q(\ell < m, \ell > c_q)
\end{eqnarray}}}
where $P_P(\ell < m, \ell > c_p)$ and $P_Q(\ell < m, \ell > c_q)$ are 
understood to be $P(\ell < m, \ell > c)$ from the single-hub-with-costs case 
  for probabilities $p$ and $q$ respectively. 
Substituting Eq. (\ref{p-ell-lt-m}) into the first term of
Eq. (\ref{thd-with-costs-before-sum}) and performing the summation yields:
{\small
\begin{eqnarray}
   P_P(\ell < m, \ell > c_p)\bigl[ 1- \sum_{i-c_q=1}^{\ell-c_q-1}P_Q(i < m, i > c_q)\bigr]
 \ \ \ \ \ \ \ \nonumber \\
\ \ \ \ \ \ \ = (g_{0pq} + g_{1pq}\ell + g_{2pq}\ell ^2)(a_pa_q)^{\ell -1}
\end{eqnarray}
}
where
{\small
\begin{eqnarray}
  a_p &=& 1-p \nonumber \\
  a_q &=& 1-q \nonumber \\
  g_{0pq} &=& (1-p)^{-c_p}(1-q)^{-c_q}p^2c_p((c_q+1)q-1) \nonumber \\
  g_{1pq} &=& (1-p)^{-c_p}(1-q)^{-c_q}p^2(1-(c_p+c_q+1)q) \nonumber \\
  g_{2pq} &=& (1-p)^{-c_p}(1-q)^{-c_q}p^2q \ \ .\nonumber
\end{eqnarray}
}
An equivalent substitution and summation performed on the second term in 
Eq. (\ref{thd-with-costs-before-sum}) yields the
same answer but with labels $p$ and $q$ interchanged.
The third term, after substitution and summation, yields:
{\small
\begin{eqnarray}
 P_P(\ell < m, \ell > c_p)P_Q(\ell < m, \ell > c_q) \ \ \ \ \ \ \ \nonumber \\
\ \ \ \ \ \ \ = (h_0+h_1\ell +h_2\ell ^2)(a_pa_q)^{\ell -1}
\end{eqnarray}
}
where
{\small
\begin{eqnarray}
  h_0 &=& (1-p)^{-c_p}(1-q)^{-c_q}p^2q^2c_pc_q \nonumber \\
  h_1 &=& -(1-p)^{-c_p}(1-q)^{-c_q}p^2q^2(c_p+c_q) \nonumber \\
  h_2 &=& (1-p)^{-c_p}(1-q)^{-c_q}p^2q^2 \ \ .\nonumber
\end{eqnarray}}
Substitution of these individual terms into Eq. (\ref{thd-with-costs-before-sum})
yields:
{\small
\begin{equation}\label{plm-cost-thd}
    P(\ell < m)= (g'_0+g'_1\ell +g'_2\ell ^2)(a_pa_q)^{\ell -1}
\end{equation}
}
where $g'_i = g_{ipq}+g_{iqp}-h_i$.
To calculate the full probability distribution for the case $\ell > c_p \geq
c_q$ we now only require $P(\ell = m)$:
{\small
\begin{equation}\label{pem-cost-thd}
    P(\ell = m) = 1-\sum_{i=c_q+1}^{c_p}P_Q(i < m)-\sum_{i=c_p+1}^{\ell-1}P(i < m)
\end{equation}
}
where $P_Q(i < m)$ is the single-hub-plus-costs distribution for a hub with
probability $q$ and $P(i < m)$ is given by Eq. (\ref{plm-cost-thd}).  We define the following
functions: 
{\small
\begin{eqnarray}
  f_x(a,n) &=& \sum_{i=1}^{n-1}i^x a^{i-1} \nonumber \\
    \tilde{f}_x(a,n_1,n_2) &=& f_x(a,n_1)-f_x(a,n_2) \nonumber \label{summation-fn1-fn2}\ \ .
\end{eqnarray}
}
We then substitute $P_Q(i < m)$ and $P(i < m)$ into Eq. (\ref{pem-cost-thd})
yielding:
{\small
{\setlength\arraycolsep{1pt}
\begin{eqnarray}
  P(\ell = m,\ell > c_p) = 1 - \frac{q^2}{(1-q)^{c_q}}\bigl[\tilde{f}_1(a_q,c_p + 1, c_q + 1) 
\nonumber \\
         -c_q\tilde{f}_0(a_q,c_p + 1, c_q + 1)\bigr]
         -\bigl[g'_0\tilde{f}_0(a_pa_q, \ell , c_p + 1) \nonumber \\
         +g'_1\tilde{f}_1(a_qa_p, \ell , c_p + 1)
         +g'_2\tilde{f}_2(a_pa_q, \ell , c_p + 1)\bigr]\ \ .
\end{eqnarray}
}
}
We now obtain the final distribution by performing the sum over $m$:
{\small
{\setlength\arraycolsep{1pt}
\begin{equation}
  P(\ell, \ell  \leq c_q) = \frac{1}{n-1}
\end{equation}
\begin{eqnarray}
  P(\ell,c_q<\ell \leq c_p) = \frac{1}{n-1}\bigl[1+(\ell -c_q-1)q \nonumber \\
    +(n-1-\ell )(\ell -c_q)q^2 \bigr](1-q)^{\ell -c_q-1}
\end{eqnarray}
\begin{eqnarray}
  P(\ell,c_p<\ell) = \frac{1}{n-1}\biggl[1-\frac{q^2}{(1-q)^{c_q}}\bigl[\tilde{f}_1(a_q,c_p +
1, c_q + 1) \nonumber \\
     -c_q\tilde{f}_0(a_q,c_p + 1, c_q + 1)\bigr]
     -\bigl[g'_0\tilde{f}_0(a_pa_q, \ell , c_p + 1) \nonumber \\
     +g'_1\tilde{f}_1(a_qa_p, \ell , c_p + 1) 
     +g'_2\tilde{f}_2(a_pa_q, \ell , c_p + 1)\bigr] \nonumber \\
     +\bigl[(n-1-\ell)(g'_0+g'_1\ell +g'_2\ell ^2) 
     (a_pa_q)^{\ell-1}\bigr]\biggr] \ \ \ \ .
\end{eqnarray}
}
}
The resulting distribution, which has an
interesting multi-modal form, is plotted in Fig.  3 for the directed case: $Q$
now depends on five variables due to the additional probability $q$ and cost
$c_q$, such that $\rho_p \equiv pn$, $\rho_q \equiv qn$, $\gamma_p \equiv
\frac{c_p}{n}$, $\gamma_q \equiv \frac{c_q}{n}$ with $z$ as before.
Interestingly if the value of $\rho_q$ increases above $\rho_p$ the distribution
tends to the single hub case extremely quickly - i.e. the P-hub is then barely
used.  If the P-hub has a high degree and a high cost, then the distribution
behaves as though the P-hub is not there until $\ell > \gamma_p$, where it
quickly falls to zero.  The undirected case is similar to the directed case since 
once again the same scaling relationship exists between them.

In summary, we have presented analytic results for a simple yet realistic model of a
congested network. Elsewhere we will discuss embedding our $N$-hub cluster within larger and
more complex networks, and will present a quantitative comparison to the transport
routings observed within laboratory-grown fungi in an attempt to understand `costs' within biological
networks. 

N.F.J. is grateful to P.M. Hui and F.J. Rodriguez for discussions, and thanks
M. Tlalka, S.C.
Watkinson, P.R. Darrah and M.D. Fricker for permission to use the image in Fig. 1(b).

\newpage


\begin{thebibliography}{99}

\bibitem{watts98} D.J. Watts and S.H. Strogatz, Nature {\bf 393}, 440 (1998).

\bibitem{nets} D. S. Callaway, M. E. J.  Newman, S. H.  Strogatz, and D. J.
Watts, Phys. Rev. Lett. {\bf 85}, 5468 (2000).

\bibitem{nets2} R. Albert and A.L.  Barabasi, Phys. Rev. Lett. {\bf 85}, 5234
(2000).

\bibitem{charges} L.A. Brunstein, S.V. Buldyrev, R. Cohen, S. Havlin and H.E.
Stanley, Phys. Rev. Lett. {\bf 91}, 168701 (2003).

\bibitem{central} 
R. Guimera, A. Diaz-Guilera et al., Phys. Rev. Lett. 89,
248701 (2002).

\bibitem{search} V. Colizza, J. R. Banavar et al., Phys. Rev. Lett. 92, 
198701 (2004).

\bibitem{gradient} Z. Toroczkai, K. E. Bassler, Nature 428, 716 (2004).

\bibitem{appl} M. Tlalka, D. Hensman, P.R. Darrah, S.C. Watkinson and M.D.
Fricker, New Phytologist {\bf 158}, 325 (2003).

\bibitem{DM} S.N. Dorogovtsev and J.F.F. Mendes, Europhys. Lett. {\bf 50}, 1
(2000).

\bibitem{railway} P. Sen, S. Dasgupta, A. Chatterjee, P. A. Sreeram,
G. Mukherjee and S. S. Manna, Phys. Rev. E {\bf 67}, 036106 (2003).

\end{thebibliography}
\end{document}